\documentclass[10pt,conference]{IEEEtran}
\usepackage{cite}
\usepackage{array,multirow}
\usepackage{graphicx}
\usepackage{hyperref}
\usepackage{balance}
\usepackage{booktabs}

\def\BibTeX{{\rm B\kern-.05em{\sc i\kern-.025em b}\kern-.08em
    T\kern-.1667em\lower.7ex\hbox{E}\kern-.125emX}}

\begin{document}

\title{Who Does What? Work Division and Allocation Strategies of Computer Science Student Teams}

\author{\IEEEauthorblockN{Anna van der Meulen}
\IEEEauthorblockA{Leiden Institute of Advanced Computer Science \\
University of Leiden, The Netherlands \\
a.n.van.der.meulen@liacs.leidenuniv.nl}
\and
\IEEEauthorblockN{Efthimia Aivaloglou}
\IEEEauthorblockA{Leiden Institute of Advanced Computer Science \\
University of Leiden, The Netherlands \\
Open University of The Netherlands \\
e.aivaloglou@liacs.leidenuniv.nl}
}

\maketitle

\begin{abstract}
Collaboration skills are important for future software engineers. In computer science education, these skills are often practiced through group assignments, where students develop software collaboratively. The approach that students take in these assignments varies widely, but often involves a division of labour. It can then be argued whether collaboration still takes place. The discipline of computing education is especially interesting in this context, because some of its specific features (such as the variation in entry skill level and the use of source code repositories as collaboration platforms) are likely to influence the approach taken within groupwork. The aim of this research is to gain insight into the work division and allocation strategies applied by computer science students during group assignments. To this end, we interviewed twenty students of four universities. The thematic analysis shows that students tend to divide up the workload to enable working independently, with pair programming and code reviews being often employed. Motivated primarily by grade and efficiency factors, students choose and allocate tasks primarily based on their prior expertise and preferences. Based on our findings, we argue that the setup of group assignments can limit students' motivation for practicing new software engineering skills, and that interventions are needed towards encouraging experimentation and learning.
\end{abstract}


\section{Introduction}
Courses in computer science curricula often involve software engineering projects that are assigned to groups of students. Group assignments are commonly the first experiences that computer science students gain in developing software collaboratively. Through group assignments, students get the opportunity to work on software projects that are of larger scale than individual course projects can be. At the same time, group projects enable students to practice their collaboration skills \cite{pfaff2003}, which are important in the software development industry \cite{Hewner_Guzdial_2010, Li_Ko_2015, Scaffidi_2018}, with current industrial trends promoting cooperative working techniques such as shared code ownership and pair programming \cite{Tirronen_2011}. However, even though collaboration and teamwork skills are important for the next generation of software engineers, it has been found that communication and teamwork skills are areas where graduating computer science students frequently fall short of the expectations and work requirements of industry \cite{Radermacher_2013, Craig_2018}.

Research in the area of group assignments has highlighted their advantages and disadvantages related to labour market preparation \cite{Feichtner1984,labeouf2014,labeouf2016, bentley2013, hattum2013}. Researched topics include the formation and set up of groups, differences in contributions between team members, and grading  \cite{labeouf2014,labeouf2016, bentley2013, hattum2013}. Further, group assignments are also used as an instructional strategy in the form of team-based learning \cite{Michaelsen_2008} and collaborative learning \cite{slavin1983, kirschner2018}. Their documented benefits concern the learning process itself, with actively working together towards a mutual learning goal \cite{teasley1993} having been found to be more effective, compared to individual approaches, in certain types of learning \cite{kirschner2009,kirschner2018}. This can be understood from the cognitive load theory, which describes the limitations of the individual working memory during complex tasks and the benefits of sharing this task in a group \cite{kirschner2009}. Together, these different research lines form the basis of understanding the set up, experiences, and outcomes of students' group assignments. 



Groupwork in the field of computing education has particular characteristics that can influence how students approach and experience their group assignments. These characteristics include the fact that students enter computer science courses with varying levels of prior programming experience \cite{Wilcox, Alvarado}, that they could be given the opportunity to apply practices like pair programming in their assignments and, finally, that within software projects source code repositories can be used as collaboration platforms, making process data \cite{gousios_2008} available. The way in which students approach their group assignment and the workload, which often involves some division of labour \cite{paulus2005, sormunen2014, saleh2011}, might be affected by these particular aspects of computing education.

The aim of this study is to examine the following research question: \emph{What are the work division and allocation strategies that university-level computer science students employ during their group assignments?} This is approached from the students' perspective of these strategies, in order to gain an in-depth understanding of their experiences and motivations. To answer our research question, we interviewed 20 final-year Bachelor's and Master's students from four research-intensive universities in the Netherlands about their experiences and perceptions on group programming assignments throughout their studies.

Our thematic analysis revealed that students tend to initially divide up the workload so that they can work independently, while commonly employing practices such as code reviews and pair programming. Their motivation while dividing and distributing the workload is most often the grade outcome and efficiency, and rarely the potential learning benefits. The allocation of tasks to group members is commonly guided by the preferences, skills, and expertise that the members already possess, with students often identifying specific tasks (for example, front-end development) that they prefer to take upon.
 
\section{Background and Related Work}

\subsection{Collaboration Strategies}
The overall approach of students to groupwork in educational settings can vary significantly \cite{paulus2005}. Division of labour has been found to be a common starting point \cite{paulus2005, sormunen2014, saleh2011}. This entails that, at the start of an assignment, students divide up the work into separate tasks, and assign a task and related responsibility to each group member who works on and completes this task individually \cite{paulus2005, saleh2011}. Different ways of working, some of which entail division of labour, have also been identified \cite{paulus2005, sormunen2014}. These include pair collaborations (where two members of the group work together on an activity), group collaborations (where multiple members of the group work together on an activity), and delegation (involving one individual taking sole responsibility, for instance, for an overall check of the end product) \cite{sormunen2014}. Studies identifying these approaches come from various fields and topics, including writing assignments \cite{sormunen2014} and engineering projects \cite{saleh2011}.

From the students' perspective, an important consideration in deciding which approach to take, and favoring the division of labour, is `efficiency of work progress’ \cite{sormunen2014}. The reasoning behind dividing up the work is that it allows students to focus on or specialize on their specific task, and, combined with delegation, choosing which students are most suited for which aspect. Students can however also take into account `quality of the work process', which can result in them choosing strategies of pair or group collaboration \cite{sormunen2014}. This difference in prioritizing work efficiency or process is interesting from the perspective of educational aims within groupwork. Previously the distinction between `cooperation' and `collaboration' has been made \cite{paulus2005}. Cooperation specifically refers to the strategy of dividing up a group task and having the resulting parts completed entirely individually by the group members \cite{henri1996}. Collaboration, in turn, explicitly involves a continued process to \textit{construct and maintain a shared concept of a problem} \cite{roschelle1995}. Consequently, which approaches are taken by students, and whether a certain approach is desirable to be stimulated, are important from the perspective of the aim of a group assignment as well. Furthermore, the starting point of division of labour is likely to be highly prevalent in computing education, because of the different levels of prior knowledge and expertise that computing students can have \cite{Wilcox, Alvarado}.
 
\subsection{Computer Science Work Division and Allocation Strategies}
Within the approach to divide up an overall assignment into sub tasks and responsibilities, an important step consists of the allocation of these different parts of the work to the group members. Insights on the allocation of work mainly come from research on professional software development teams. The issue of task allocation has received special attention within the context of distributed software development \cite{Filho}, where it is recognized as a major challenge due to an insufficient understanding of the criteria that influence task allocation decisions \cite{Lamersdorf}.
When introducing self-organizing teams in agile software development, the most important barrier has been found to be the developers' highly specialized skills and the corresponding division of work \cite{Moe2008}. At the same time, expertise coordination has been found to be crucial among software development teams, since it strongly affects team performance \cite{Faraj}. The performance of software teams was also found to be positively impacted by knowledge diversity and a proper level of task conflict, indicating disagreement among team members regarding the content of the tasks
being performed \cite{Liang2007}. Lin et al. quantitatively analyzed the effects of team member's competence and task difficulty on their workload variation \cite{JunTA}, while Amrit examined the effect of social network structures on task distribution \cite{Amrit}. Specific team roles in software teams (for example, team leader, systems analyst, programmer) have been linked to personality traits \cite{Gorla, Ferreira}. Overall, it appears that, in the professional field of software development, specialization of skills and diversity of knowledge are important factors in teamwork, which can both be challenging and potentially beneficial.

Examining groupwork in computing education, Lingard and Berry analyzed data from 39 teams working on software projects and found a significant correlation between team project success, team synergy and the degree to which work is equitably shared among team members \cite{Lingard} without, however, examining specific work distribution strategies. Process data can give insight into work distribution. Automated tools have been proposed for monitoring student collaboration and contributions, including tools utilizing students' wikis and software version control system (svn) repositories \cite{Kim_assistinginstructional}, analyzing online team discussion transcripts to visualize team mood, role distribution and emotional climate \cite{ Tarmazdi_2015}, and calculating productivity metrics describing student contributions in their git repositories \cite{janjaap}.

\section{Methodology}
\label{sec:setup}

The goal of this study is to explore the strategies that university-level computer science students employ to approach the workload in group programming assignments. Our focus is on their work division and task allocation strategies. We aim to gain insights on the extent to which collaboration occurs and the extend to which new knowledge is gained and practiced. To this end, 20 semi-structured interviews were conducted with students from computer science departments of four universities in the Netherlands. The following paragraphs describe the participants and the interview and data analysis process.

\subsection{Participants}
Twenty computer science students participated in the research. Participants were invited during their classes and lab sessions, as well as through Slack and other student communication channels. Students who were at a later stage of their study, i.e. finally year Bachelor (twelve participants) or Master (eight participants), were recruited in order to capture a wide range of experiences with group programming assignments. Further, students were recruited from four different public research-intensive universities in the Netherlands, including a university of technology and a university that offers distance learning opportunities. After 20 interviews, it was determined that no new information was being gathered and thus saturation was reached.

At least four students from each university were included. Four of the Master's students had completed their Bachelor degrees in other universities, three of which in other countries. The participants indicated to originate from (alphabetically): Bangladesh, India, the Netherlands, Switzerland, Ukraine and the U.S. Gender was reported by all participants (5 female and 15 male), age by most (known age range was between 20 and 33), and some participants reported to have autism or a speech disorder. The self-reported expected grade of the participants for the degree that they were pursuing varied from 6,5 to 9, with a mean value of 7,5 out of 10.

\subsection{Interview Process}
The current research question on students' collaboration and work division strategies was part of a broader project on students' experiences with group programming assignments. The interview protocol consisted of 11 questions in five topics: background, experiences with group programming assignments, perceptions on assignment setup, experiences with grading, and perceptions on grading. In order to answer the current research question, information from the background and the first two topics was used. The interview questions for these topics can be found in Table 1. 

\begin {table*}
\caption {Interview Questions}
\begin{center}
\begin{tabular}{|l|l|l|}
\hline
Topic & Questions \\
\hline
\multirow{2}{8em}{Background} & Can you tell something about your personal background, where are you from/how old are? \\
& Can you tell something about your studies and programming background? \\
\hline
Programming experience & What is your experience with group programming assignments?\\
\hline
\multirow{4}{8em}{Perceptions on assignment set up} & Assume that you have a group programming assignment to do with another one or two team members. \\
& How do you decide what to do? \\
& How do you collaborate within the team? For example, do you program together, do you give feedback? \\
& What do you think of how this team work goes and your own role in it?\\
\hline
\end{tabular}
\end{center}
\end{table*}

After the first two interviews, with one male and one female student, the scope and questions of the interview protocol were reconsidered, after which they remained unchanged for the other interviews. Ten of the interviews were conducted face-to-face and the other ten through Skype. Of all sessions voice recordings were made, which were transcribed with automatic transcription software. A manual check and correction of the transcriptions was performed after this. The length of the interviews ranged from 12 to 58 minutes, with an average length of 23 minutes.

\subsection{Data Processing}
The methodological framework followed in processing the data was a thematic analysis approach \cite{braun2006}. In this approach, patterns (referred to as themes) are identified in qualitative data in order to organize and described the data and, further, interpret them in the context of the research topic. Thematic analysis is flexible in that the determination of themes can be both theory- and data-driven \cite{blair2015,braun2006}. For the current research, the main themes were theory driven, while subthemes (referred to below as labels) within these main themes were datadriven, generated from the interview data. This application of a thematic analysis approach is fitting since we build upon existing knowledge on collaboration strategies, yet openly examine how this takes form in the specific population of computer science students. The process is described in detail below. 

First, seven themes were determined based on the research questions of the project, closely in line with the topics of the interview protocol: student profile, experience with group assignments, work division and collaboration strategy and motivation, task allocation strategy, organisation of the group work, experience with grading, and perception on grading. Next, labels were developed within these themes. Initially, the two researchers independently labelled two interviews. For example, within the theme ``work selection strategy and motivation" the label 'how to divide/approach work division' was generated. The labels were compared and, together with the themes, discussed by the researchers, after which the theme ``experience with group assignments" was added. Both researchers continued to independently label four additional interviews to continue generating appropriate labels. After discussing together again, the researchers determined fixed labels within each theme. 

Second, all interviews were coded using these labels. For the current study, only the themes student profile, experience with group assignments, work division and collaboration strategy and motivation, and task allocation strategy were further processed. The interviews were divided between the researchers. Doubts as to which label was appropriate for a given excerpt were discussed. 

Third, for each theme the information was integrated across participants, as well as when applicable integrated across themes. For example, information on the profile of the students (institution and current degree) was included in the description of the other themes.
    

\section{Results}
\label{sec:results}
The participants have been assigned random numbers and are referred to as S1 to S20. The quotes that are included in the description are verbatim. To protect the anonymity of the participants, identifiable information has been suppressed and person pronouns have been converted to the male form. The results are analyzed under two main parts in relation to the themes: (1) work division strategy, and (2) task allocation strategy.

In terms of experience with group programming assignments, almost all students (18) responded that they have participated in several group assignments during their studies as part of several courses. Group size varies; students refer both to groups of two people and, commonly, of larger size. Two undergraduate students from the same university reported having none to very limited (only one group assignment with one other person) experience.

\subsection{Work Division Strategy}
All students who have participated in group assignments described work division strategies. Table 2 presents an overview of the work division strategies grouped under three sub-themes:  project startup, collaboration after initial division, and motivation factors. Even though this study is not a quantitative study, Table 2 includes the frequency of repetition of each label, to allow identification of the most commonly mentioned labels. 

\begin {table*}
\caption {Theme 1: Work Division and Collaboration Strategies in Group Programming Assignments} 
\begin{center}
\begin{tabular}{|l|l|l| } 
\hline
Sub-themes & Categories & F \\
\hline
\multirow{4}{8em}{Project startup} & Dividing up assignment into sub-components or tasks \& everyone continues by themselves & 17 \\ 
& Agreeing about the content of the assignment together & 8 \\ 
& Looking at the qualities and experiences of all group members & 5 \\ 
& Directly dividing up the work without additional discussions & 4 \\ 
\hline
\multirow{6}{8em}{Collaboration after initial division} & Pair programming  & 12 \\ 
& Code reviews  & 10 \\ 
& Regular check-in/contact & 8 \\ 
& Contact only for questions and problems & 3 \\ 
& Contact only for integrating at the end & 1 \\
& Work separately on the same parts & 1\\
\hline
\multirow{6}{8em}{Motivation for work division and collaboration strategy} & Good/sufficient grade  & 10 \\ 
& Less work for everyone  & 7 \\  
& Time management & 6 \\  
& Learning benefits  & 3 \\ 
& Clear work division & 3 \\ 
& Not possible to work on the same code & 2 \\ 
\hline
\multicolumn{2}{l}{F: Frequency of repetition}
\end{tabular}
\end{center}
\end{table*}

\subsubsection{Project startup}
All students who have participated in group assignments describe the process of getting started with the assignments. The same main strategy was described by all but one of these students: \emph{dividing up the the overall assignment into sub-components or tasks that everyone can continue to work on by themselves}. The student who does not refer to this strategy describes the process of working together in a pair. 

Surrounding this main process of dividing op the overall work and having group members continue on their own, different additional strategies were described. Explaining how they start up the project, eight students talk about \emph{looking at and agreeing about the content of the assignment together}, by going through the work, asking each other questions about it, making an overview, agreeing on the scope or end product, or, as one participant describes, even starting with creating the basics together. For instance: 
\begin{itemize}
    \item ``What we do is first thing we just looked through it, we take like one day maybe to all read the assignment and ask questions if something is not clear. But when everybody reads the assignment and everything is clear, we, the first thing we do usually, is just try to split it into parts" (S8), and 
    \item ``So we've all made a list of topics and then discuss which ones everybody thought for themselves. And then we pick, we had discussed them and picked one and that one we worked on'' (S1).
\end{itemize}

Five other students talk about getting started by \emph{looking at the qualities and experiences of all group members}, and four participants only mention the start of the process as \textit{directly dividing up the work without additional discussions or considerations}. One student describes the experience of different ways to get started with the group assignment, either explicitly defining roles and tasks or ``it just all clicked" (S4) and everyone started working. 

\subsubsection{Collaboration after initial division}
After the start of the project, there are a variety of ways (mentioned by 12 students) through which group members keep in touch, re-group, and in general collaborate once everyone starts working on their individual parts. There is quite some difference in how extensive this contact appears to be, ranging from regularly discussing and keeping each other updated to only asking questions when needed. Three approaches appear to be (1) a \textit{regular check-in}, for instance in weekly meetings, to discuss and possibly set new tasks, and/or to be in touch when needed, (2) three students mention only being in touch \textit{in case of a question or problem}, and (3) one says only \textit{at the end to try to put things together}. 

Ten students also mention the aspect of \textit{reviewing each other’s code}. One student indicates that they rarely do any code reviews, and two participants say they do it to a limited extent, checking the code briefly or only when there is problem. The majority, seven students, talk about more extensive peer review, with two even indicating that they set rules or make agreements about checking each other’s code, for instance always having someone double check another person’s code:
\begin{itemize}
\item ``And our rule is that when someone has dragged it to the check column, someone else has to check on Github whether the code is actually correct and they drag it to the done column. So if everything goes well, everything will be checked before it's actually considered done'' (S17).
\end{itemize}

One student indicates that the importance of code review is stressed by the teacher, whereas three students say that they themselves want to review the code to make sure that they understand everything, and because (according to one) this improves the quality or (according to another) it is necessary to make sure the code can be connected. As one student describes:
\begin{itemize}
\item ``So that's how we felt like to, you know, all have a mutual understanding of what the code of the project is, that we should review each other's code and also of course for quality'' (S5).
\end{itemize}

Some students also reflect on checking each other’s work and code. Two students indicate, within their view on code reviews, that they do trust the other team members. Two other students describe the process of distinguishing between giving suggestions and pointing out actual errors, which they do, and really perfecting the algorithm or implementing changes in other’s work, which they don’t. As one of these students indicates: 

\begin{itemize}
\item ``So in the end, it's also you, you look at each other's work and you give suggestions. You can propose your ideas, you can propose some changes, but you never really make those changes yourself, even though you know, you can have a better, sometimes I know I have a better idea'' (S8).
\end{itemize}

Two other, specific ways of working are also brought forward. The first is \textit{pair programming}, mentioned by 12 students. Six of these students indicate that they do pair programming within group assignments, two of whom only occasionally, which depends on whether there are exercises in the assignment for which this is a good fit, whether others also prefer this way of working, and whether its works out practically. Two students explicitly describe their reasons for preferring pair programming: you can ask questions, share your ideas, check with each other, and it is more creative. As one students says: 

\begin{itemize}
\item ``I prefer pair programming, so you can like share your ideas and like the logic and someone can check it also like whether you are doing it correctly or not'' (S6).
\end{itemize}

Five other students indicate that they do not engage a lot or at all in pair programming, either because that is not common in their university, it is not common within group assignments, time constraints make it difficult, or because they do not prefer it. One student only engages specifically in pair debugging.  One other student indicates that there are advantages in terms of learning in pair programming only for the one of the pair who has a lower skill level: 

\begin{itemize}
\item ``Even though you might learn a lot [...] or you can give someone a lot. But for someone that knows a lot, uh, it's a time waste because how does he benefit from it?'' (S18).
\end{itemize}

The second other way of working is only described by one student, and concerns an approach within group assignments where both group members \emph{work separately on the same parts}, to later compare and then take the one that is best. 

\subsubsection{Motivation for work division and collaboration strategy}
Several motives are described for the selection of work division and collaboration strategies. Most students explain that they engage in the approach of splitting up the work because it \emph{helps them get a good or sufficient grade}. For instance:
\begin{itemize}
    \item ``We both wanted to have a high grade and use the techniques that were discussed in the literature, but we didn't know exactly what the capabilities of the other were. So, someone can say, I can do this, but you don't know whether it is according your own standards. So that's why we had to find out, what's your level of experience? What's the way you code? [...] So you get to know each other and, and then make a decision how to make the best of it'' (S16), and
    \item ``The end goal is usually to get the highest grade possible. So that's different from learning the most possible'' (S17).
    
\end{itemize}

Work division is also preferred because it involves \textit{less work for everyone} and is necessary in terms of \textit{time management}, it has \textit{learning benefits}, it is a fair approach where it is \textit{clear what everyone has to do}, or because it is \textit{not possible/useful} to both work on the same code. Reasons to get started together  include that it is important to make sure everyone is on the same process, to understand and sketch out the assignment together, and to see what everyone’s talents are. It is mentioned by two students that how to divide up the work depends on the specific assignments.

\subsection{Task Allocation Strategy}
Most (17) students who have participated in group assignments described task allocation strategies. Table 3 presents an overview, grouped under two sub-themes: the assignment of specific tasks, and the assignment of team roles.

\begin {table*}
\caption {Theme 2: Task Allocation Strategies in Group Programming Assignments} 
\begin{center}
\begin{tabular}{|l|l|l| } 
\hline
Sub-themes & Categories & F \\
\hline
\multirow{6}{8em}{Assignment of specific tasks to group members} & Preferences of each team member & 9 \\ 
& Member skills, experiences, what they're good at, task familiarity & 8 \\ 
& Content of particular sub-components (front/back-end, etc.) & 7 \\ 
& Effect of skills acquired during the studies & 3 \\ 
& No specific factors & 3 \\
& Being social / female & 1 \\
\hline
\multirow{2}{8em}{Assignment of roles in the team} & One person taking in a leadership role  & 7 \\
& Staying in the background and having others make the decisions  & 1 \\ 
\hline
\multicolumn{2}{l}{F: Frequency of repetition}
\end{tabular}
\end{center}
\end{table*}

\subsubsection{Assignment of specific tasks}
Continuing on the process of dividing up a group assignment into sub-components or tasks, almost all (17) students discuss the assignment of specific tasks to group members. Often mentioned is to consider the skills, experience, or interests that group members have, and to divide up and assign tasks accordingly. Specifically, eight students indicate that parts are assigned based on the \emph{team member skills, experiences, what they good at}, or, as one student indicates, what is familiar to them. One student describes that this is difficult to determine at first, and that it is necessary to adjust along the way:
\begin{itemize}
    \item ``The sort of like the amount of progress any single person can make. Um, and that's just sort of like, well, how did you split it up? Is this person better at this kind of problem or this kind of problem? Or maybe this person is just much more productive or just much better at programming or much smarter'' (S7).
\end{itemize}

Further, nine students in total mention considering the \emph{preferences of each team member}:
\begin{itemize}
\item ``It was basically just you do what you feel, um, you should do'' (S1), and 
\item ``Since I have some prior expertise or like I feel like taking this part'' (S10). 
\end{itemize}

Interestingly, three students describe how the \emph{skills developed through their studies affected the tasks they would take upon}. Two of these students indicate that, at the beginning of their studies, they were less experienced and had to take on the easier parts, whereas, as one of them specifically describes, later on they could take on more:
\begin{itemize}
    \item ``Um, in the earlier projects you, they're quite simple and uh, there's always someone who is a lot better than the rest. So you can give those people who are fairly new to programming the more learning-based tasks'' (S4), and
    \item `Well as I experienced it, like in the beginning, uh, the first year it was like I wasn't very good at programming. So, uh, it was very hard with, uh, I did like a small percentage of the tasks but, as I got better, I did higher percentages and like, uh, did more work for the programming assignments'' (S13).
\end{itemize}

The other student refers more to a general development of people becoming experts and specialization becoming possible: 

\begin{itemize}
\item ``So in the early group projects, um, so early in the Bachelor's you have to do certain things, like everybody has to program this or that. Um, but as you go later in the Bachelor, you get more freedom. So the more freedom we got, the more we decided to really let the person do what they're very good at to, to work as efficiently as possible'' (S5). 
\end{itemize}

Apart from mentioning expertise or preference at a general level, several students also describe other or more detailed motivations to choose or to be assigned specific sub-components of the assignment. Seven students refer to the \emph{content of particular sub-components}. The students talk about specific types of activities (often mentioning both what they do like and what they don’t like), including user interface and visual aspects versus algorithmic problems, back-end programming, background logistics, web development, trying out different algorithms, designing the structure versus programming, or more abstract aspects, such as work on the overview and the logistics behind the project. For example,
\begin{itemize}
\item ``I usually work with friends who love these algorithmic problems and then they take them and then I'll take, you know, the user interface, the visual things, graphical work, mathematical work'' (S3), and
\item `I usually leave the heavy back-end programming for the people I know that are better at that then than I am. So I was very soon in, like early in my bachelor's, I recognized that that is not my best part in programming. Um, so I always try to leave that to people who are better at it than I and take the parts that I felt I'm good at and that I can work with'' (S5).
\end{itemize}

One student has experienced that certain parts are expected of her because of her \emph{being social or being the only girl} in the group:

\begin{itemize}
\item ``I always ended up doing the presentation, either because I look relatively social or because I was the girl in the group or whatever'' (S2).
\end{itemize}

Finally, three students do \emph{not specify concrete factors for their choices}, either specifying they really don't have any true interest, that they just pick what needs to be done first, or that it depends on the project. 

\subsubsection{Assignment of roles}
Eight students talk about the roles that are assigned within group projects, referring not to tasks or activities but to the place members have within the group. All but one student talk specifically about \emph{one person taking in a leadership role}:

\begin{itemize}
\item ``Within all the other groups, it was always someone who took the lead and that doesn't necessarily mean uh, content wise, but maybe just communicating with the teacher, making sure everybody's there on time'' (S1), and
\item ``And you can all be in a group, you can all be equals, but at some point you, you'll find someone who will take on some kind of leader form. But if you don't have that person, the group can become kind of unguided'' (S2).
\end{itemize}

One student who does not refer to someone taking in a leadership role only mentions himself preferring to \emph{stay at the background and having others make the decisions}. Out of the other seven, all but one indicate to have taken in this leadership role themselves. The other student has not experienced anyone taking this role, but thinks it is something that should happen and that should be implemented as part of the assignment:

\begin{itemize}
\item ``So either there should be a system like that where you need to pick who is responsible for the whole group, who is responsible for particular parts of a group or the teacher could assign specific, like who could break down the assignments and confirm who is going to do which part'' (S9).
\end{itemize}

For four students, taking in the role of leader is a common and deliberate choice: 

\begin{itemize}
\item ``I like to have a general overview of the project. Right. So usually connecting everything and uh, UI related things'' (S5), and
\item ``So, most of the time I start the project, um, and lead and we will just decide who is best in what part and we'll really split it up. And if someone is having a hard time, we will meet up and sit together. That's what I like to do best'' (S19).
\end{itemize}

Two other students have only taken in the leadership role incidentally, which seems to depend on the specific group.

\section{Discussion}
The research question of this study was: what are the work division and allocation strategies that university-level computer science students employ during their group assignments? Overall, our findings indicate that students tend to divide up the work, and choose and assign tasks primarily based on their preferences and prior expertise. At the same time, several joined practices are mentioned, such as brainstorming sessions at the start, regularly checking in, and adjusting tasks and responsibilities along the way. The motivations of the students were found to mainly include grading and efficiency but also, less commonly, wanting to learn and having a fair approach for all group members.

\subsection{Dividing and Allocating Based on Preferences and Prior Expertise}
The computer science students participating in this research appear to be well aware of the approach they take in their group assignments and what they take into consideration. Clearly, division of labour is central in their approach, in line with previous insights in education within different disciplines \cite{paulus2005, sormunen2014, saleh2011} as well as in professional software development teams \cite{Faraj, Liang2007}. The way in which the students describe their approaches also illustrates that this division of labour can take diverse forms, often involving several joined practices. These primarily take place at the start, when students discuss the scope of the assignment or get started with the basics. In addition however, some practices continue through the working process, such as checking each other's code or reconvening to discuss the progress and next steps. 

Our findings indicate that prior programming knowledge and expertise is one of the deciding factors for task allocation among student teams. This is in line with findings on the work distribution of professional software development teams, where expertise, along with the availability of people, are the most important criteria for task allocation \cite{Lamersdorf}, as well as with findings from engineering students \cite{saleh2011}. The students of this research often indicate to be motivated by specific aspects of their studies  (for example focusing on user interface and visual aspects, or focusing on algorithmic programming tasks), also suggesting the presence of early specialization. Some students also describe how their relatively low programming skills at the start of their studies impacted their role in groupwork. At the same time, however, previous research on groupwork as an instructional strategy has looked at effects of asymmetry in knowledge or expertise between members on effective collaboration \cite{kirschner2018}, substantiating that different levels of expertise and, likely related, in preference are factors that generally have a common role in groupwork approaches. 

\subsection{Presence of Collaborative Practices}
The identification of collaboration and work division strategies that students engage in, especially in the context of the specific field, is valuable in itself. However, it also gives insight into the extent to which these practices do in fact reflect collaboration. Setting a group of students together with an assignment does not automatically entail that collaboration takes place \cite{paulus2005}. Especially concerning the approach of dividing up the work, it has been questioned whether collaboration, referring to a joined thinking process, \cite{roschelle1995} occurs \cite{paulus2005}. However, although the students of this study clearly favour a set up where a significant portion of the groupwork is being done by them individually, the mix of practices they use, as well as their description of their experiences and motivations, also suggests that overall elements of collaboration appear present. For instance, brainstorming together at the start and regularly checking in does appear to be in line with a joined thinking process. There are also several practices that reflect that students take collaboration into account. As described quite insightfully by two students, they consider what the correct practice is when checking other group member's code. They will check and point out errors, but not actually do the other person's work or implement their own ideas. This could be seen as sophisticated collaborative behavior. Previously it has also been argued that collaboration and cooperation are ultimately not that distinct as they are often presented \cite{strijbos2007}. Our findings show that a hybrid approach is often applied, where parts of groupwork are completed entirely by individual members of the team, and at the same time a joined thinking process occurs to a certain extent.    

Further, the question arises of whether this approach is in line with the purpose of the students' assignments and education. It remains questionable whether the overall approach of labour division serves the learning goals of computer science study programs. If team members repeatedly opt to work on the tasks that are most familiar to them throughout the group projects within the computer science curriculum, this finding could indicate that they are led to premature specialization. This was evident in the responses of some students who revealed commonly being assigned tasks of specific nature, for example front-end development, because they were the best in their teams at these specific tasks. Mitigating this tendency to specialize on specific software development tasks might not be trivial, as long as team efficiency and performance are what motivates task allocation decisions. It might be an interesting consideration here whether such specialization is to a certain extent desired both as a natural part of their studies and for their future professional software engineering career. Ultimately, the fact that specialization of skills and diversity of knowledge is seen both as a challenge and an advantage in professional software development teams \cite{Faraj, Liang2007}, reflects that this is a larger discussion on optimization of teamwork in software development.

\subsection{Limitations}
Our research was based on interviews conducted in a small number of students of four universities of one country. The experiences and perceptions of this student population may differ from the ones of other students in the same or other institutions, countries and cultures. Additionally, students who consent to participate in interviews about group work and have their answers used in research projects may not reflect the general student population. Moreover, the reason behind the choice of recruiting final-year Bachelor's students and Master's students was to include participants with sufficient experience to describe and to inform their perceptions. This, however, left inexperienced students out of our sample. In the interviews the students often described experiences from the early years of their studies and their perceptions at the time, but their reflections might have been influenced by the experience that they gained since then.

Regarding the internal validity of our study, a threat is the social pressure that the respondents might have felt when disclosing their perceptions about group work and about the policies they have encountered during their studies. Overall, all students appeared comfortable during their interviews and not hesitant to give their honest opinion. Still, they might have answered differently with another interviewer or in a more anonymous data collection setting. There were quite some differences between the interviews in total duration, whether they were conducted physically or online, and whether they were conducted in the native language of the students. All students did however seem at ease and fluent in English, and variation in the interview duration is not uncommon in a semi-structured approach. Concerning data processing, in the case of a thematic analysis of the type of data as included in the current research, decisions on the approach are guided by the underlying aim of the study \cite{braun2006}. In the case of our study, the aim was to explore the perception of students, therefore, within our pre-specified topics, the approach was data-driven. The different experiences and ideas of the students were integrated yet described extensively, giving context and providing quotes to illustrate and substantiate our interpretation.

\section{Concluding Remarks}
The aim of this study was to explore the experiences and perceptions of computer science on their work division and allocation strategies within group assignments. The use of semi-structured interviews proved valuable, since it showed students' underlying motivations and reasoning within the overall favored approach for division of labour. The hybrid approach that computer science students appear to take, in which mostly individual completions of tasks are combined with several practices (including brainstorming sessions at the start, regularly checking in, and adjusting tasks and responsibilities along the way) that suggest a joined thinking process, provides an understanding of the way in which groupwork is approached in the specific field of computer science. Moreover, these findings show that it is important to consider what the educational aims of group assignments are in the first place, and how these aims can best be fostered in the set up and instruction of the assignments. 

Our results suggest several possible directions for future work. The effect of the varying skills and prior programming experience, as well as of other possible characteristics such as gender, could be studied in depth through both qualitative and quantitative studies. A larger scale quantitative study could also assess whether and how the factors identified through our interviews are interrelated. It is important that future research on groupwork concerning programming assignments takes into account the aspects where this type of groupwork is similar or deviates from group work in other disciplines and topics. Further, research on how the educational aims of group assignments in computing education can be fostered should include measures and interventions for encouraging students to practice new software engineering skills and take upon tasks that they are not already familiar with. Finally, the relation to future work roles and expectations within software engineering teams could be researched further.

\section*{Data availability}
In order to protect the privacy of the participants in our study and due to potentially identifiable information in the interview transcripts, the data collected in this research are not made publicly available.

\section*{Acknowledgments}
The authors would like to thank Marina Milo (Vrije Universiteit Amsterdam) for helping with setting up the interviews and processing the interview transcripts. We would also like to thank the students that were involved in this research. We are grateful for your time and for sharing your experiences and insights with us.

\bibliographystyle{IEEEtran}
\balance
\bibliography{IEEEabrv, refs}

\end{document}